\documentclass[onecolumn,nofootinbib]{revtex4}
\usepackage{graphicx} % Required for inserting images
\usepackage{setspace}
\usepackage{amsmath,amssymb}
\usepackage{graphicx}
\usepackage{dcolumn}
\usepackage{bm,color}
\usepackage{hyperref}
\usepackage{accents}
\usepackage{amssymb,float}
\usepackage{amsmath}
\usepackage{multirow}
\usepackage{tikz}
\usepackage[rightcaption]{sidecap}
\sidecaptionvpos{figure}{t}
\usepackage{enumerate}

\newcolumntype{P}[1]{>{\centering\arraybackslash}p{#1}}
\newcolumntype{M}[1]{>{\centering\arraybackslash}m{#1}}
\hypersetup{
    colorlinks=true,
    linkcolor=red,
    filecolor=magenta,      
    citecolor=blue
}
\newcommand{\udt}[3]{#1^{#2}_{\phantom{#2}#3}}

\newcommand{\dut}[3]{#1_{#2}^{\phantom{#2}#3}}

\newcommand{\lc}[1]{\accentset{\circ}{#1}}%Levi-Civita connection
\newcommand{\dd}{\mathrm{d}}

\begin{document}

\title{Stable bouncing solutions in Teleparallel Horndeski gravity:\\ violations of the no-go theorem}

\author{Bobomurat Ahmedov}
\email{ahmedov@astrin.uz}
\affiliation{Institute of Fundamental and Applied Research,
National Research University TIIAME, Kori Niyoziy 39, Tashkent 100000, Uzbekistan}
\affiliation{Institute of Theoretical Physics, National University of Uzbekistan, Tashkent 100174, Uzbekistan}

\author{Konstantinos F. Dialektopoulos}
\email{kdialekt@gmail.com}
\affiliation{Department of Mathematics and Computer Science, Transilvania University of Brasov, 500091, Brasov, Romania}

\author{Jackson Levi Said}
\email{jackson.said@um.edu.mt}
\affiliation{Institute of Space Sciences and Astronomy, University of Malta, Malta, MSD 2080}
\affiliation{Department of Physics, University of Malta, Malta}

\author{Abdurakhmon Nosirov}
\email{abdurahmonnosirov000203@gmail.com}
\affiliation{New Uzbekistan University, Movarounnahr street 1, Tashkent 100000, Uzbekistan}
\affiliation{Ulugh Beg Astronomical Institute, Astronomy St.  33, Tashkent 100052, Uzbekistan}

\author{Zinovia Oikonomopoulou}
\email{zhnobia.oikonomopoulou.21@um.edu.mt}
\affiliation{Institute of Space Sciences and Astronomy, University of Malta, Malta, MSD 2080}

\author{Odil Yunusov}
\email{odilbekhamroev@gmail.com}
\affiliation{New Uzbekistan University, Movarounnahr street 1, Tashkent 100000, Uzbekistan}
\affiliation{Ulugh Beg Astronomical Institute, Astronomy St.  33, Tashkent 100052, Uzbekistan}

\date{\today}

\begin{abstract}
In order to have singularity-free solutions at the beginning of the Universe, we need to violate the null energy condition. In the general class of Horndeski gravity, there are healthy NEC-violating solutions, which however are plagued with instabilities or some kind of pathologies in the history of the Universe; this is known as the no-go theorem. In this paper, we study the possibility of stable bouncing solutions in the Teleparallel analog of Horndeski gravity and we show explicitly that there exist healthy violations of the no-go theorem. 
\end{abstract}

\maketitle

\section{Introduction}

New physics in the early Universe has become one of the strongest candidates for alleviating some of the most pressing problems in modern cosmology. Recent measurements of the accelerating expansion of the Universe \cite{SupernovaSearchTeam:1998fmf,SupernovaCosmologyProject:1998vns} indicate that it is expanding at a faster rate than predicted by the $\Lambda$CDM concordance model \cite{DiValentino:2021izs}. This has led to a reevaluation of potential departures from the $\Lambda$CDM concordance model, and opened the possibility of a resolution to problem of the as yet unobserved cold dark matter (CDM) \cite{Baudis:2016qwx,Bertone:2004pz,Gaitskell:2004gd}, as well as the theoretically problematic cosmological constant \cite{Peebles:2002gy,Copeland:2006wr}. The latest problems in the concordance model primarily come from the Hubble tension \cite{DiValentino:2020zio}, but the growth of the large scale structure of the Universe has also shown signs of problematic features $f\sigma_{8,0}$ \cite{DiValentino:2020vvd}. The collective problem of cosmological tensions has been reinterpreted as a problem between direct measurements in the late Universe \cite{Riess:2021jrx,Wong:2019kwg,Anderson:2023aga,Freedman:2020dne}, and inferred cosmological parameter values based on early time observations \cite{Aghanim:2018eyx,DES:2021wwk,eBOSS:2020yzd,Zhang:2021yna,Cooke:2017cwo}. 

The early Universe offers a rich area in which to propose exotic physics to confront problematic elements in the late Universe. One prime example is the introduction of inflation \cite{Guth:1980zm,Linde:1981mu} to the concordance model, which has been shown to resolve the horizon and flatness problems as well as other seemingly insurmountable obstacles. The very early Universe is again being investigated for possible alternative scenarios such as cyclic and bouncing solutions which can produce nonsingular cosmologies. For general relativity (GR) minimally coupled with an energy-momentum tensor $T_{\mu\nu}$ this issue turns out to be challenging, as expressed through the Penrose theorem \cite{PhysRevLett.14.57}. The heart of the problem comes from the null energy condition (NEC) where
\begin{equation}
    T_{\mu\nu} k^{\mu} k^{\nu} \geq 0\,,
\end{equation}
for every null vector $k^{\mu}$. For a flat cosmological background, the direct consequence is that an isotropic and homogeneous Universe leads to a Hubble parameter that decreases backward in time, thus assuring that a big bang singularity must have taken place. Mild alternatives to GR such as the addition of a canonical scalar field produce a generically satisfied NEC $ T_{\mu\nu} k^{\mu} k^{\nu} = \dot{\phi}^2 \geq 0$, while a non-canonical field that also depends on the kinetic term can violate the NEC but pose a significant level of instability in the cosmologies they produce \cite{Armendariz-Picon:1999hyi,Garriga:1999vw}.

The full spectrum of models that contain a single scalar field and its kinetic term while adhering to second order equations of motion are contained in the general class of Horndeski theory models \cite{Horndeski:1974wa}. Horndeski theory, and its subclasses, have garnered a lot of interest in recent decades since they consolidate a vast swath of models in a relatively manageable form. However, the recent multimessenger observations of gravitational waves together with their electromagnetic counterpart has severely constrained the breadth of models that are observational compatible unless a frequency-dependency is added to the class of models \cite{TheLIGOScientific:2017qsa,Ezquiaga:2017ekz}. Regardless of this, the full class of Horndeski models features a question on its stability when probed for non-singular initial cosmological models \cite{Kobayashi:2016xpl}. While many subclasses of Horndeksi gravity exist in which the early Universe is stable and the NEC is circumvented, these models invariably turn out to host instabilities in other parts of the cosmology \cite{Cai:2012va,Kobayashi:2015gga,Rubakov:2014jja}. This has been termed the so-called no-go theorem and ranges from the sound speed evolving into negative values, or other types of singularities appearing in the evolution of cosmic expansion, among others.

The compounding problems of foundational instability and the severe constraints imposed on Horndeski theory due to multimessenger observations have led to interest in the literature on other approaches to producing general classes of cosmological models. One example is beyond Horndeski theory \cite{Kobayashi:2019hrl,Traykova:2019oyx} where higher order terms are admitted to the traditional form of Horndeski theory with the limitation that they do not admit Ostrogradsky instabilities in the ensuing theory. While interesting, this generalization seems to not circumvent the underlying problem in the speed of propagation of gravitational waves. Another approach that has gathered interest in the literature is to revisit the underlying foundations of Horndeski gravity and to exchange the geometric curvature of the model with teleparallel torsion \cite{Bahamonde:2021gfp,Cai:2015emx,Krssak:2018ywd}. Here the Levi-Civita connection $\udt{\lc{\Gamma}}{\sigma}{\mu\nu}$ (over-circles refers to the quantities calculated with the regular Levi-Civita connection) is replaced by the teleparallel connection $\udt{\Gamma}{\sigma}{\mu\nu}$. A formulation of this exists in which the gravitational action can be made equal (up to a total divergence term) to the Einstein-Hilbert action, called the teleparallel equivalent of general relativity (TEGR) \cite{Hehl:1994ue,Aldrovandi:2013wha}. This is dynamically equivalent to GR admitting identical solutions. On the other hand, teleparallel gravity (TG) has led to a plethora of new physics models ranging from traditional approaches to modified gravity such as f(T) gravity \cite{Ferraro:2006jd,Ferraro:2008ey,Bengochea:2008gz,Linder:2010py,Chen:2010va,Bahamonde:2019zea, RezaeiAkbarieh:2018ijw,Farrugia:2016pjh,Cai:2015emx}, as well as less traditional manifestations of gravitational physics such as the generalization of Horndeski gravity.

The way that TEGR is constructed as the active Einstein-Hilbert component together with a total divergence, or boundary, term shows how TG is generically lower order in nature as compared with curvature-based gravitational theories. When constructing a generalized model in which TEGR is modified by the inclusion of a single scalar field and its kinetic term, this led to the same terms as appear in traditional Horndeski gravity as well as a new term which has had important phenomenological consequences \cite{Bahamonde:2019shr}. Termed Bahamonde-Dialektopoulos-Levi Said (BDLS) theory, this formulation of TEGR with a generalized scalar field and kinetic term produces a speed of gravitational waves that revives the more exotic components of regular Horndeski gravity \cite{Bahamonde:2019ipm}, while also producing a rich landscape of gravitational wave polarizations \cite{Bahamonde:2021dqn}. The parameterized post-Newtonian parameters also do not place heavy constraints on the model space of the theory \cite{Bahamonde:2020cfv}, and ghost and Laplacian stability analyses also do not seem to overly limit the possible models allowable in the theory \cite{Capozziello:2023foy}. A number of nontrivial BDLS models have been recently suggested in the literature, based on an analysis of Noether symmetries \cite{Dialektopoulos:2021ryi}, as well as on conditions for well-tempered cosmologies \cite{Bernardo:2021bsg,Bernardo:2021izq}.

The recent presentation of the full cosmological perturbative analysis has opened the way for a more robust analysis of the cosmological evolution of BDLS theory \cite{Ahmedov:2023num}. The no-go theorem has been shown to hold for the entire class of Horndeski theory models \cite{Kobayashi:2016xpl}. In this work, we analyze possible violations of this theorem, and thus, potential cosmological models in which healthy resolutions to the initial singularity problem are found. We limit our analysis to cases where physical models are evolved giving reasonable evolution profiles for cosmological parameters are found. We thus aim to show a series of examples in which the no-go theorem is circumvented in a healthy way. In Sec.~\ref{sec:BDLS_intro}, we introduce the expression of BDLS theory and the no-go theorem formally. We then discuss our main results in Sec.~\ref{sec:examples} where we discuss the general issue of the no-go theorem in BDLS theory. Finally, we close with a summary and some conclusions in Sec.~\ref{sec:conclusion}.

\section{The no-go theorem}\label{sec:BDLS_intro}

We consider the BDLS action \cite{Bahamonde:2019shr}, which is the teleparallel analog of Horndeski gravity, which reads
\begin{equation}\label{action}
    \mathcal{S}_{\text{BDLS}} \sim \int d^4 x\, e\mathcal{L}_{\text{Tele}} + \sum_{i=2}^{5} \int d^4 x\, e\mathcal{L}_i+ \int d^4x \, e\mathcal{L}_{\rm m}\,,
\end{equation}
where $e$ is the determinant of the tetrad. The third term on the right hand side is a regular matter Lagrangian. The second one is the sum of all the regular Horndeski contributions~\cite{Horndeski:1974wa}, expressed in the teleparallel geometry, i.e.  
\begin{align}
    \mathcal{L}_{2} & :=G_{2}(\phi,X)\,,\label{eq:LagrHorn1}\\[4pt]
    \mathcal{L}_{3} & :=-G_{3}(\phi,X)\lc{\square}\phi\,,\\[4pt]
    \mathcal{L}_{4} & :=G_{4}(\phi,X)\left(-T+B\right)+G_{4,X}(\phi,X)\left[\left(\lc{\square}\phi\right)^{2}-\phi_{;\mu\nu}\phi^{;\mu\nu}\right]\,,\\[4pt]
    \mathcal{L}_{5} & :=G_{5}(\phi,X)\lc{G}_{\mu\nu}\phi^{;\mu\nu}-\frac{1}{6}G_{5,X}(\phi,X)\left[\left(\lc{\square}\phi\right)^{3}+2\dut{\phi}{;\mu}{\nu}\dut{\phi}{;\nu}{\alpha}\dut{\phi}{;\alpha}{\mu}-3\phi_{;\mu\nu}\phi^{;\mu\nu}\,\lc{\square}\phi\right]\,.\label{eq:LagrHorn5}
\end{align}
The circle on top of a quantity denotes its calculation with the Levi-Civita instead of the teleparallel connection. One notices that the only difference compared to the regular Horndeski Lagrangian is that terms like the Ricci scalar are now expressed with its teleparallel analog,
\begin{equation}
    \lc{R} = -T + B\,,
\end{equation}
and $\lc{G} _{\mu\nu}$ is similarly the Einstein tensor. $T$ is the torsion scalar, defined from the torsion tensor, i.e. $T ^A{}_{\mu\nu} = 2 \Gamma ^A{} _{[\mu \nu ]}$. and $B$ is a boundary term. In addition, exactly as in Horndeski gravity, we have four arbitrary functions of the scalar field $\phi$ and its kinetic term $X = - (\partial ^\mu \phi \partial _\mu \phi )/2$, i.e. the  $G_i$'s with $i = 2,3,4,5$. What is more, the first term in the action~\eqref{action} is a new arbitrary function of the form
\begin{equation}
\label{eq:LTele}
    \mathcal{L}_{\text{Tele}}:= G_{\text{Tele}}\left(\phi,X,T,T_{\text{ax}},T_{\text{vec}},I_2,J_1,J_3,J_5,J_6,J_8,J_{10}\right)\,,
\end{equation}
where the arguments are described in Table~\ref{tab:scalars}. The torsion tensor can be decomposed into three irreducible parts
\begin{align}
    a_\mu &:= \frac{1}{6}\epsilon _{\mu\nu\lambda\rho}T^{\nu\lambda\rho} \,,\\
    v_\mu &:= T^\lambda {}_{\lambda\mu}\,,\\
    t_{\lambda \mu\nu} &:= \frac{1}{2} \left(T_{\lambda \mu\nu }+ T_{\mu\lambda \nu}\right) + \frac{1}{6}\left( g_{\nu\lambda} v_{\mu} + g_{\nu\mu} v_{\lambda} - \frac{1}{3}g_{\lambda \mu}v_\nu \right)\,, 
\end{align}
which are the purely axial, vector and tensor parts respectively and $\epsilon _{\mu\nu\lambda\rho}$ is the totally antisymmetric Levi-Civita tensor in four dimensions. Based on this decomposition we can define the following scalar invariants
\begin{equation}
    T_{\rm ax} := a_\mu a^\mu\,,\quad T_{\rm vec} := v_\mu v^\mu\,,\quad T_{\rm ten} := t_{\lambda\mu\nu}t^{\lambda \mu\nu}\,.
\end{equation}
\begin{table}[H]
\centering
\begin{tabular}{|c|c||c|c|} 
\hline
$T$  & $\frac{3}{2}T_{\rm ax} + \frac{2}{3}T_{\rm vec} - \frac{2}{3}T_{\rm ten}$ & $J_3$  & $v_\sigma t^{\sigma \mu\nu} \phi _{;\mu} \phi _{;\nu}$ \\
\hline
$T_{\rm ax}$  & $a_\mu a^\mu$ & $J_5$  & $t^{\sigma\mu\nu}t_\sigma{}^\beta{}_\nu \phi _{;\mu}\phi _{;\beta}$ \\
\hline
$T_{\rm vec}$  & $v^\mu v_\mu$ & $J_6$  & $t^{\sigma\mu\nu}t_{\sigma\alpha\beta} \phi _{;\mu}\phi _{;\nu}\phi^{;\alpha}\phi^{;\beta}$ \\
\hline
$I_2$  & $v^\mu \phi_{;\mu}$ & $J_8$  & $t^{\sigma\mu\nu}t_{\sigma\mu\alpha} \phi_{;\nu}\phi ^{;\alpha}$ \\
\hline
$J_1$ & $a^\mu a^\nu \phi _{;\mu} \phi _{;\nu}$ & $J_{10}$  & $ \epsilon ^\mu {}_{\nu\rho\sigma} a^\nu t^{\alpha\rho\sigma}\phi_{;\mu}\phi_{;\alpha}$ \\
\hline
\end{tabular}
\caption{Description of the teleparallel scalars.}\label{tab:scalars}
\end{table}
The field equations both for the metric and for the scalar field are of second order but we do not include them here for simplicity; one can find them in \cite{Bahamonde:2021gfp,Bahamonde:2019ipm}. 
For the background, let us consider a flat FLRW tetrad of the following form, where without loss of generality the spin connection is considered to be zero
\begin{equation}\label{FLRW-tetrad}
    e^A{}_\mu = {\rm diag}(N(t),a(t),a(t),a(t))\,.
\end{equation}
The scalar field will also inherit the isometries of the tetrad in the background and thus $\phi = \phi (t)$. Linear perturbations around this background were studied in \cite{Ahmedov:2023num}. In summary, if we consider the unitary gauge where the scalar field perturbations vanish, i.e. $\delta \phi = 0$, the quadratic action for the tensor perturbations of the tetrad, $h_{ij}$, reads
\begin{equation}
   S_{\rm T}^{(2)} = \int \dd t \dd ^3 x \, \,\frac{a^3}{4} \left[\,\mathcal{G}_{\rm T} \dot{h}_{ij}^2 - \frac{\mathcal{F}_{\rm T}}{a^2}(\bm{\nabla} h_{ij})^2\,\right]\ ,
    \label{eq:ten_pert_action}
\end{equation}
while for the scalar perturbations, $\zeta$, it reads
\begin{align}
\label{quadraticscalaraction}
     S_{\rm S}^{(2)}=\int  \dd t \dd ^3 x a^3 \Bigg[\mathcal{G}_{\rm S}\dot{\zeta}^2-\frac{\mathcal{F}_{\rm S}}{a^2}(\bm{\nabla}\zeta)^2\Bigg] .
\end{align}
The coefficients are written as
\begin{gather}
    \mathcal{G}_{\rm T} = 2 \left(G_4-2 X G_{4,X}+ X G_{5,\phi}-H X \dot{\phi} G_{5,X} + 2 X G_{{\rm Tele}, J_8}+\frac{X}{2} G_{{\rm Tele}, J_5} -  G_{{\rm Tele},T}\right),
    \label{tensorcoef1} \\
    \mathcal{F}_{\rm T} = 2 \left(G_4-X G_{5,\phi}-X \ddot{\phi} G_{5,X}-G_{{\rm Tele},T} \right),
    \label{tensorcoef2}
\end{gather}
and
\begin{gather}
    \mathcal{G}_{\rm S} = 3  \mathcal{A}+\frac{\Sigma \mathcal{A}^2}{\Theta^2} \,,     
    \label{scalarcoef1}  \\ 
    \mathcal{F}_{\rm S} = \frac{1}{a}\frac{\dd}{\dd t}\Bigg(\frac{a  \mathcal{A}\mathcal{C}}{\Theta}\Bigg)-\mathcal{B} \,.\label{scalarcoef2}
\end{gather}
The functions $\Sigma$ and $\Theta$ are complicated expressions of the $G_i$'s and $G_{\rm Tele}$ and thus smooth functions of time and can be found in detail in \cite{Ahmedov:2023num}; for completeness, we also add them in the Appendix~\ref{Sec:appendix}. The rest of the functions, $\mathcal{A}, \mathcal{B}$ and $\mathcal{C}$ are expressed as
\begin{align}
    \mathcal{A}&=\mathcal{G}_{\rm T}+f_1\left(G_{\rm Tele}\right)\,, \label{function-A} \\
    \mathcal{B}&=\mathcal{F}_{\rm T}+ f_2\left(G_{\rm Tele}\right)\,, 
     \label{function-B} \\
    \mathcal{C}&=\mathcal{G}_{\rm T}+f_3\left(G_{\rm Tele}\right)\,.\label{function-C}
\end{align}
where the analytic expressions for $f_i(G_{\rm Tele})$ with $i = (1,2,3)$ are given in the Appendix~\ref{Sec:appendix}.

Clearly, from \eqref{eq:ten_pert_action} and \eqref{quadraticscalaraction}, we see that in order to avoid ghost instabilities we require that 
\begin{equation}\label{stability-ghost}
    \mathcal{F}_{\rm T} > 0, \quad \mathcal{G}_{\rm T} > 0,
\end{equation}
while in order to avoid gradient instabilities we require that
\begin{equation}\label{stability-grad}
    \mathcal{F}_{\rm S} > 0, \quad
    \mathcal{G}_{\rm S} > 0.
\end{equation}
Notice that in contrast to regular Horndeski, even in the minimally coupled case where $G_4 =$ const, $G_5 = 0$ and $G_{{\rm Tele},\phi} = 0 = G_{{\rm Tele},X}$, $\mathcal{F}_{\rm T}$ and $\mathcal{G}_{\rm T}$ are still time dependent, because of their dependence on non-trivial terms of $G_{\rm Tele}$. Also, the speed of the gravitational waves is given by the ratio $c_{\rm T}^2 = \mathcal{F}_{\rm T}/\mathcal{G}_{\rm T}$ and it can still be equal to unity, even when $G_{4,X} \neq 0$ and $G_{5,\phi} \neq 0$ (see \cite{Bahamonde:2019ipm} for more details); while the sound speed of the scalar perturbations is given by $c_{\rm S}^2 = \mathcal{F}_{\rm S}/\mathcal{G}_{\rm S}$.

The coefficient $\mathcal{F}_{\rm S}$ from Eqs.~\eqref{scalarcoef2},\eqref{function-B} and \eqref{function-C} can be written as
\begin{equation}
    \mathcal{F}_{\rm S} =\frac{1}{a}\frac{\dd}{\dd t}\Bigg(\frac{a\mathcal{A}\mathcal{G}_{\rm T}}{\Theta}\Bigg)-\mathcal{F}_{\rm T}+\frac{1}{a}\frac{\dd}{\dd t}\Bigg(\frac{a\mathcal{A}f_3(G_{\rm Tele})}{\Theta}\Bigg)-f_2(G_{\rm Tele}) = \frac{1}{a}\frac{\dd \xi}{\dd t} - \mathcal{F}_{\rm T} + \mathcal{F}_0\,,
\end{equation}
where we have defined the quantities
\begin{equation}\label{eq:xi}
     \xi := \frac{a\mathcal{A}\mathcal{G}_{\rm T}}{\Theta}  \quad {\rm and}\quad \mathcal{F}_0 := \frac{1}{a}\frac{\dd}{\dd t}\Bigg(\frac{a\mathcal{A}f_3(G_{\rm Tele})}{\Theta}\Bigg)-f_2(G_{\rm Tele}) \,.
\end{equation}
The absence of gradient instabilities \cite{Capozziello:2023foy} leads to 
\begin{gather}
    \mathcal{G}_{\rm S}>0 \Rightarrow \mathcal{A} \left(3+ \frac{\Sigma \mathcal{A}}{\Theta ^2}\right)>0 \Rightarrow \mathcal{A} \neq 0 \,,\\
    \mathcal{F}_{\rm S}>0 \Rightarrow \frac{\dd \xi}{\dd t} > a (\mathcal{F}_{\rm T}-\mathcal{F}_0) > 0\,.\label{eq:grad_insta_cond}
\end{gather}
Note that $\xi$ can vanish only if $a = 0$ (singularity), because $\Theta$, being a function of the $G_i$'s and $G_{\rm Tele}$, as well as $\phi$ and $H$, it is supposed to be a smooth function of time and thus finite everywhere. Thus, integrating Eq.~\eqref{eq:grad_insta_cond}, from some $t_i$ to $t_f$, we get
\begin{equation}\label{key}
    \xi_f - \xi_i > \int_{t_i}^{t_f}a\left(\mathcal{F}_{\rm T}-\mathcal{F}_0\right) \dd t\,.
\end{equation}

Let's pause here for a bit to consider the regular Horndeski theory, where $G_{\rm Tele} = 0$ and thus $\mathcal{F} _0 = 0$. The equation \eqref{key} was used in \cite{Libanov:2016kfc,Kobayashi:2016xpl} to prove the no-go theorem in a sub-class and in the full Horndeski theory, respectively. 

Summarizing their results, if we consider a non-singular, expanding universe, i.e. $a>{\rm const.}>0$ for $t\rightarrow - \infty$, the integral in the right hand side of \eqref{key} from $t_i \rightarrow - \infty$ to $t_f \rightarrow + \infty$ may be convergent or not depending on the asymptotic behaviour of $\mathcal{F}_{\rm T}.$ Specifically, if $\mathcal{F}_{\rm T}$ approaches zero sufficiently fast at $\pm \infty$, the integral does converge; however, this would indicate potential strong coupling in the tensor sector and that is why we avoid it. We thus assume that the integral does not converge and $\xi _i <0,$ then Eq.~\eqref{key} becomes
\begin{equation}
    - \xi _f < |\xi_i| - \int _{t_i} ^{t_f} a \mathcal{F}_{\rm T}\dd t\,.
\end{equation}
Since $a >0$ and $\mathcal{F}_{\rm T}>0$ for increasing $t$, the integral becomes positive, the right hand side becomes negative and thus $\xi _f > 0$, which means that $\xi$ crosses zero, which is impossible for any $t$ in a non-singular universe. The same argument holds if we assume that $\xi >0$ everywhere, and thus non-singular models are pathological and the no-go theorem is proven. 

Coming back to our case and the teleparallel analog of Horndeski gravity, the presence of $\mathcal{F}_0$ in Eq.~\eqref{key} indicates the possibility of stable non-singular solutions in the context of this theory. As we will show, this is indeed the case. Let us assume that $\xi_i<0$; in this case we can write Eq.~\eqref{key} like 
\begin{align}\label{assume1}
    -\xi_f<|\xi_i|-\int_{t_i}^{t_f}a\left(\mathcal{F}_{\rm T}-\mathcal{F}_0\right) \dd t.
\end{align}
If the integral is an increasing function, then the right hand side of inequality~\eqref{assume1} is negative as $t$ increases and $\xi_f>0$; in which case $\xi$ crosses zero and the no-go theorem is extended in the teleparallel case as well. However, the integral in Eq.~\eqref{assume1} is not necessarily an increasing function of $t_f$ because of the presence of $\mathcal{F}_0$. In what follows, we will see specific examples in which there are stable non-singular solutions and thus the no-go theorem is violated.

\section{Evading the no-go theorem in BDLS theory}\label{sec:examples}

In this section, we will present several examples of models in the teleparallel Horndeski framework, which accept bouncing solutions that are stable and have no pathologies whatsoever. In order to find the models we follow a reconstruction method, where we assume that a bouncing solution of the form \cite{Banerjee:2019CQG}
\begin{equation}\label{eq:a-bouncing}
    a (t) = a_0 (1+ b t^2)^{1/3}\quad \text{and} \quad  \phi (t) = t \,,
\end{equation}
is a solution of the models and by assuming a general form of the $G_i$'s and $G_{\rm Tele}$ we solve the background equations for those arbitrary functions.

\subsection{Model I}\label{sec:model1}
The first model that accepts Eq.~\eqref{eq:a-bouncing} as solutions, is the one with the Horndeski functions being
\begin{align}
    G_2(\phi,X) &= X^2 \left(\frac{4 \,b}{3(1+b \phi^2)}\right)+\frac{b \, \phi}{1+b \,\phi^2} \,,\label{eq:model1_G2}\\
    G_{3}(\phi,X) &= X+\frac{1}{2} \log{(1+b\,\phi^2)} + \frac{4b\phi}{3(1+b \phi^2)}+\frac{8\sqrt{b}}{3}\tan^{-1}(\sqrt{b}\phi) \,,\label{eq:model1_G3}\\
    G_4 (\phi,X) &= 1 + m X \,,\label{eq:model1_G4} \\
    G_5 (\phi,X)&= {\rm const.}\,,
\end{align}
and the teleparallel Lagrangian will have the form
\begin{equation}
    G_{\rm Tele} = -m T X + 4 m J_5\,.\label{eq:model1_GT}
\end{equation}
$m$ is an arbitrary constant that has to be $m\geq 10$ in order for the stability criteria \eqref{stability-ghost}, \eqref{stability-grad} to be met. Notice that even though $J_5$ vanishes at the background \eqref{FLRW-tetrad}, it becomes non-trivial at the perturbative level.

\begin{figure}[!ht]
    \centering    
    \includegraphics[width=0.32\textwidth]{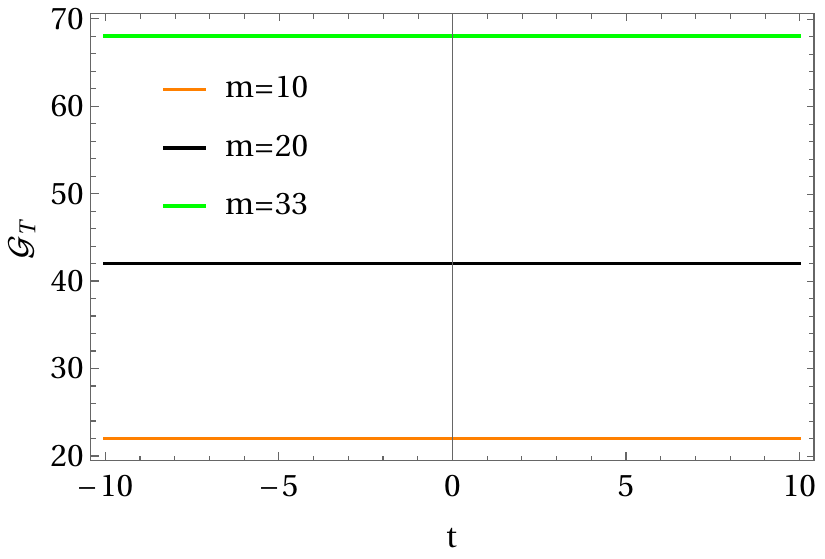}
    \includegraphics[width=0.32\textwidth]{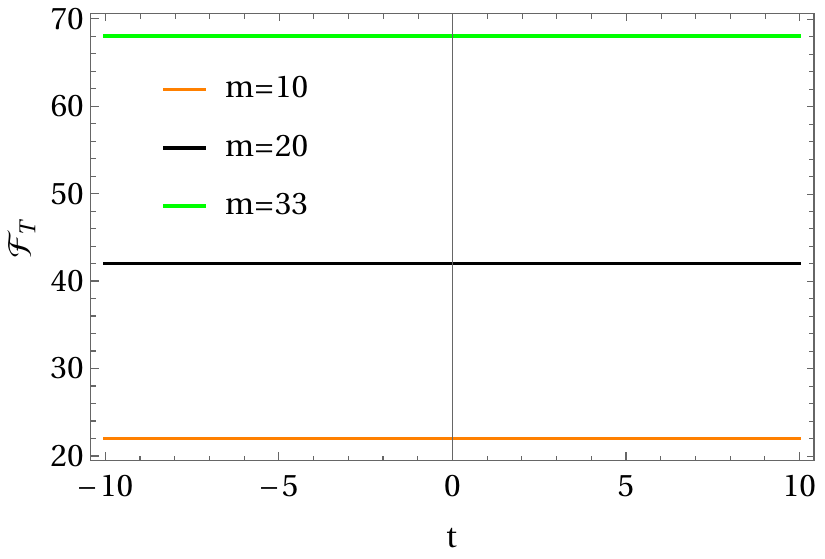}    \includegraphics[width=0.32\textwidth]{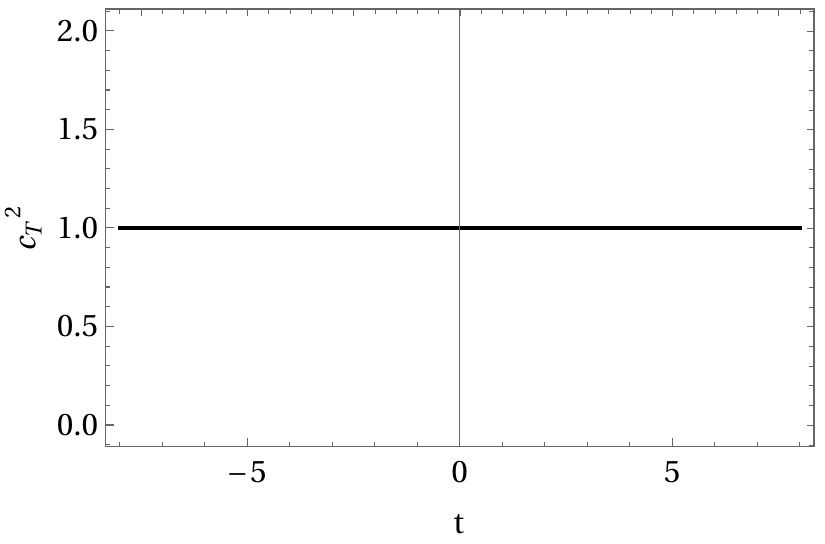} 
    \caption{Model I: The coefficients of the tensor perturbation are plotted for different values of $m$, together with the speed of gravitational waves.}
    \label{fig:model I - tensor}
\end{figure}

\begin{figure}[!ht]
    \centering
    \includegraphics[width=0.32\textwidth]{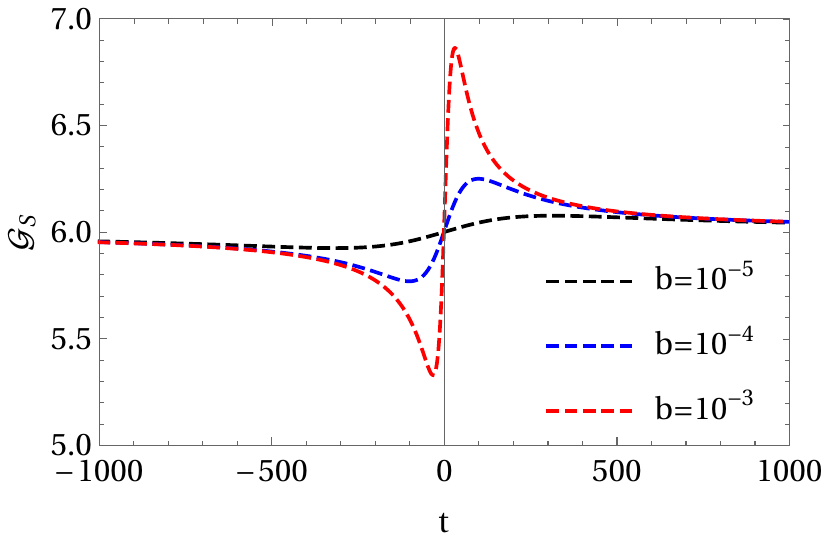}
    \includegraphics[width=0.32\textwidth]{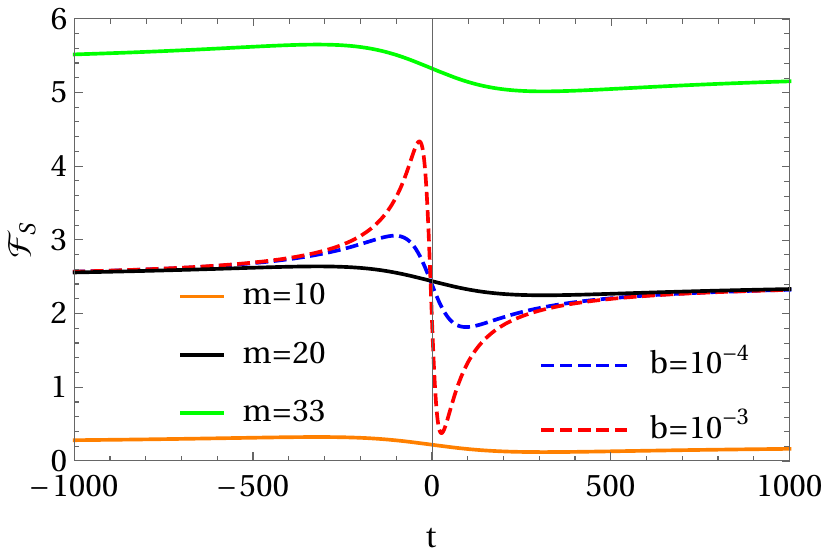}    \includegraphics[width=0.32\textwidth]{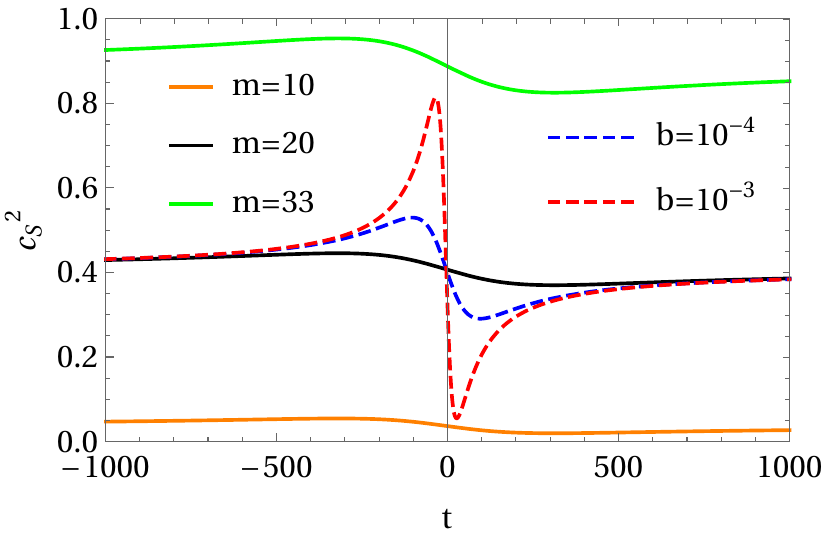} 
    \includegraphics[width=0.32\textwidth]{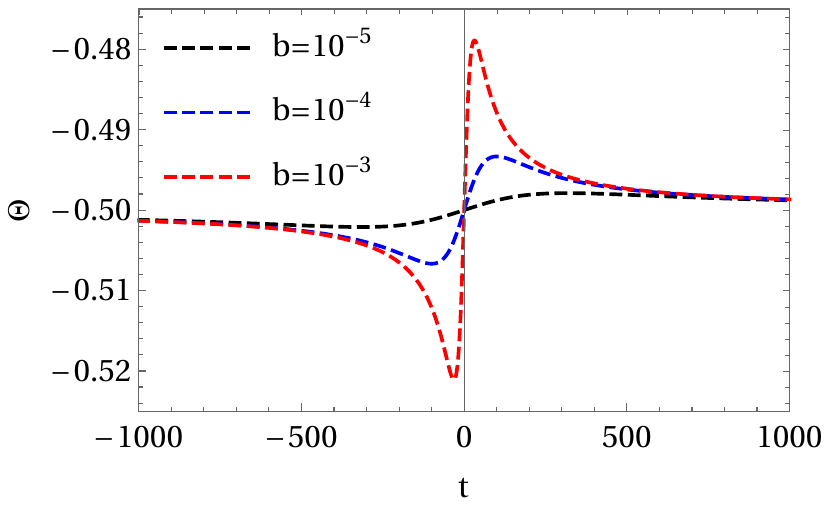}    \includegraphics[width=0.32\textwidth]{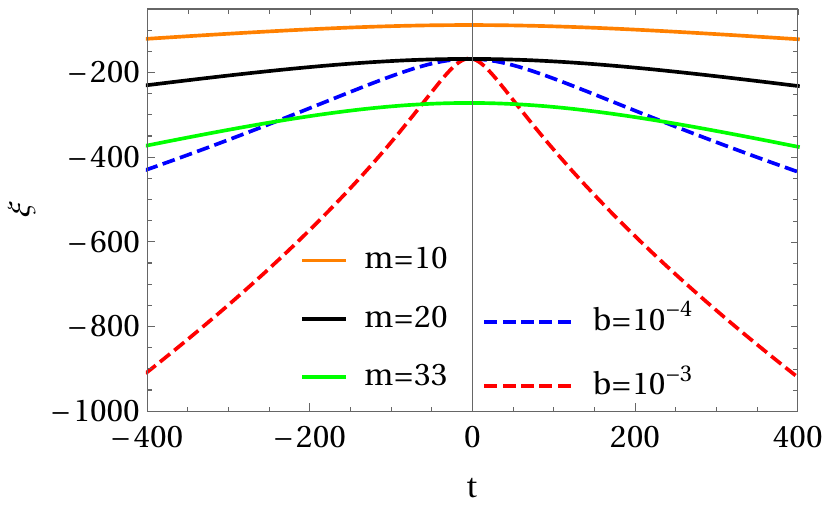}  
    \caption{Model I: The coefficients of the scalar perturbation are plotted for different values of $m$ and $b$, together with the sound speed of their propagation.}
    \label{fig:model I - scalar}
\end{figure}

For this model, the tensor perturbation coefficients \eqref{tensorcoef1},\eqref{tensorcoef2} become
\begin{equation}   
   \mathcal{G}_{\rm T} = 2 (1+m) \,, \quad  \mathcal{F}_{\rm T} = 2(1+m) \,,\\ 
\end{equation}
and thus the propagation speed of the tensor modes is always unity, i.e. $c_{\rm T}^2 = 1$. We present them for different values of $m$ in Fig.~\ref{fig:model I - tensor} 
Respectively, the scalar ones \eqref{scalarcoef1}, \eqref{scalarcoef2} take the form
\begin{gather}
    \Theta = 2H-\frac{1}{2} \,,\,\,\, 
    \mathcal{A}=2 \,,\,\,\,
    \mathcal{B}=\frac{2}{9}\Bigg(9-m\Bigg)\,,\,\,\,
    \mathcal{C}=2+\frac{20 m}{9} \, , \\ 
    \Sigma  =18H^2+6\dot{H}+\frac{9}{2}H  \, ,
\\
\mathcal{G}_{\rm S} = 6 \Bigg(1+4\frac{12H^2+4\dot{H}+3H}{(4H-1)^2}\Bigg) \,,\,\,\, \mathcal{F}_{\rm S} = 8\frac{4H^2-H+4\dot{H}}{(4H-1)^2}(1+\frac{10\,m}{9})-2\,(1-\frac{m}{9})\ , 
\end{gather}
where $H=\dot{a}/a$ is Hubble parameter. In Fig.~\ref{fig:model I - scalar} we plot them as a function of time. As one can see, for $a_0 = 1$, the value of $m$ affects the height of the plot, meaning that the sound speed could become superluminal for some values. Furthermore, the parameter $b$ distorts the actual line of the plot. For all the values considered here, i.e. $a_0 = 1$, $m= 10, 20, 33$ and $b = 10^{-i}$ with $i = 3,4,5$, the value $\Theta$ does not cross zero, meaning that $\xi$ remains finite everywhere and the stability criteria are met, i.e. both $\mathcal{G}_{\rm S}>0$ and $\mathcal{F}_{\rm S} >0$.

\subsection{Model II}

\begin{figure}
    \includegraphics[width=0.32\textwidth]{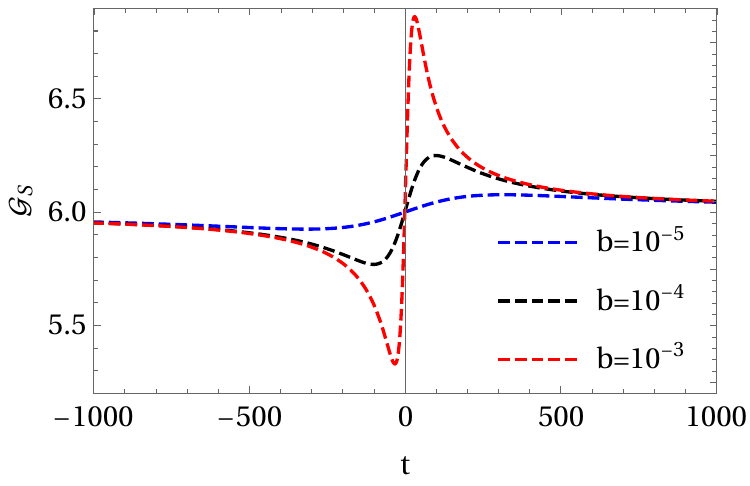}
    \includegraphics[width=0.32\textwidth]{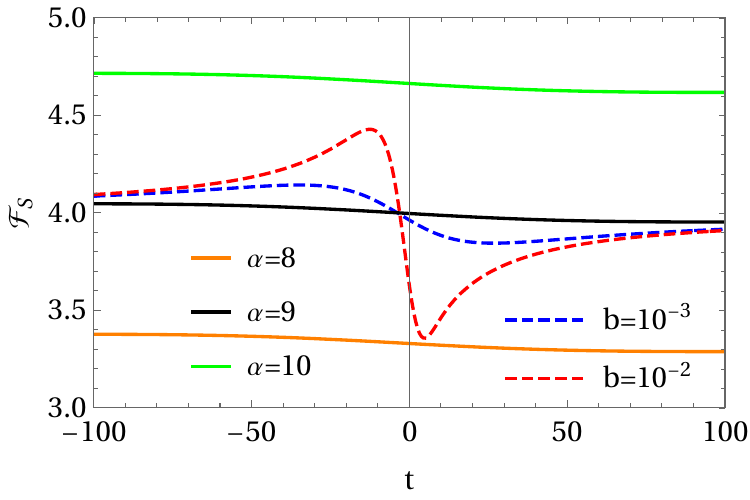}    \includegraphics[width=0.32\textwidth]{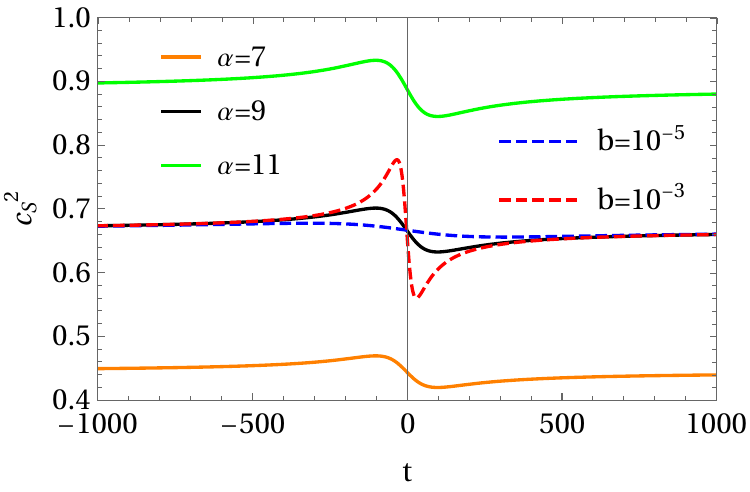} 
    \includegraphics[width=0.32\textwidth]{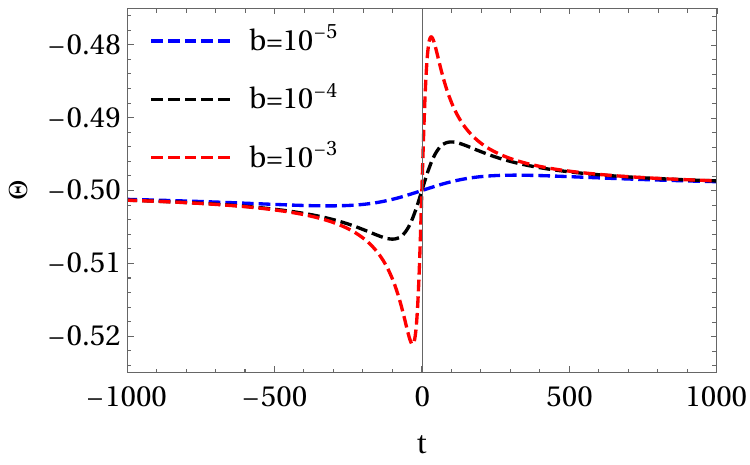}    \includegraphics[width=0.32\textwidth]{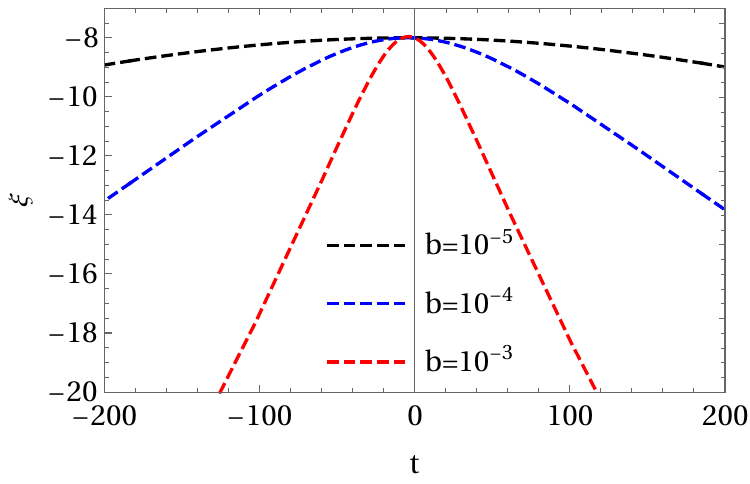}  
    \caption{Model II: The scalar perturbation coefficients are plotted for different values of $\alpha$ and $b$, together with the respective sound speed. $\Theta$ is also shown to be negative everywhere and finite.}
    \label{fig: model II - scalar}
\end{figure}

The second model that has Eq.~\eqref{eq:a-bouncing} as solutions has the Horndeski functions
\begin{align}
    G_2 &=X^2\,\Bigg(\frac{16b}{3(1+b\,\phi^2)}\Bigg)+\frac{b\,\phi}{1+b\,\phi^2} \,,\label{eq:model2_G2}\\
    G_{3}&=X+\frac{1}{2}\log{(1+b\,\phi^2)}+\frac{4b\,\phi}{3(1+b\,\phi^2)}+\frac{8\sqrt{b}}{3}\,\rm tan^{-1}{(\sqrt{b}\,\phi)} \,,\\
    G_4&=\frac{1}{2}+X \,,\\
    G_5&=\phi \,,
\end{align}
and its teleparallel Lagrangian reads
\begin{equation}
    G_{\rm Tele}=-\frac{T}{2}+\alpha\,J_3\,.\label{eq:model2_GT}\,
\end{equation}
Parameter $\alpha$ is a constant and has to be greater than three because of the stability conditions, furthermore, $b$ is the constant that enters the solution \eqref{eq:a-bouncing}. The tensor perturbation coefficients in this model take the form
\begin{align}\label{example4}
    \mathcal{G}_{\rm T}=\mathcal{F}_{\rm T}=2 \,, 
\end{align}
thus having always a constant speed of gravitational waves that is equal to one and we do not plot them for simplicity. The scalar perturbation coefficients become
\begin{gather}
    \mathcal{A} = 2\,,\,\,\,\mathcal{B}=2-\frac{2\alpha}{3} \,,\,\,\,\mathcal{C}=2+\frac{\alpha}{6} , \nonumber \\ 
    \Sigma=18\,H^2+6\,H+6\dot{H}\,,\,\,\, \Theta=2\,H-\frac{1}{2}\, , \\
    \mathcal{G}_{\rm S} = 6 \Bigg(1+4\frac{3H^2+\dot{H}+H}{(4H-1)^2}\Bigg) \,,\,\,\, \mathcal{F}_{\rm S} = 8\frac{4H^2-H+4\dot{H}}{(4H-1)^2}(1+\frac{\alpha}{12})-2\,(1-\frac{\alpha}{3})\,.
\end{gather}
In the Fig.~\ref{fig: model II - scalar} we plot $\Theta, \xi$, together with $\mathcal{G} _{\rm S}, \mathcal{F} _{\rm S}$ and the respective sound speed. As in the previous case, different values of the $\alpha$ parameter shift the height of the plot, while $b$ changes its form. For the plotted values, i.e. $a_0 = 1$, $\alpha = 7,9,11$ and $b = 10^{-i}$ with $i = 3,4,5$, the scalar perturbation coefficients are always positive, meaning that ghost and gradient instabilities are not present, while $\Theta$ does not cross zero and finite at all times.  

\subsection{Model III}

\begin{figure}
    \includegraphics[width=0.32\textwidth]{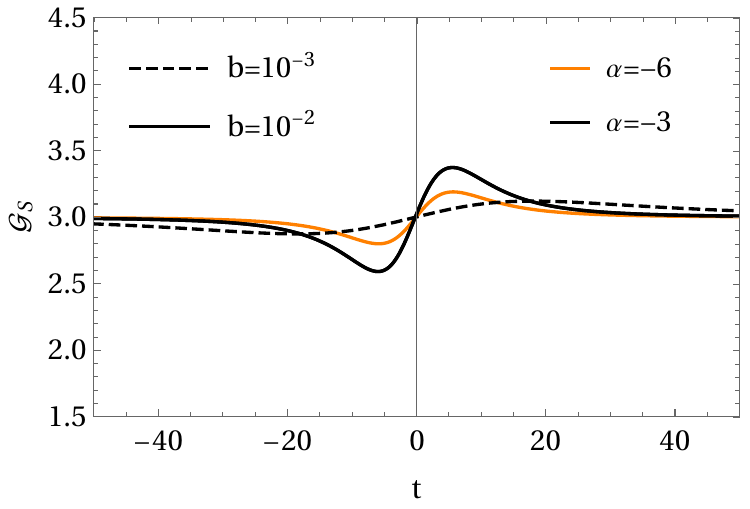}
    \includegraphics[width=0.32\textwidth]{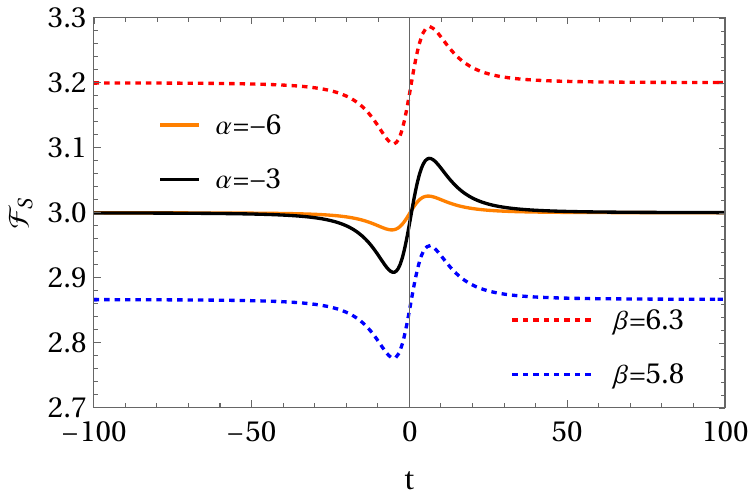}    \includegraphics[width=0.32\textwidth]{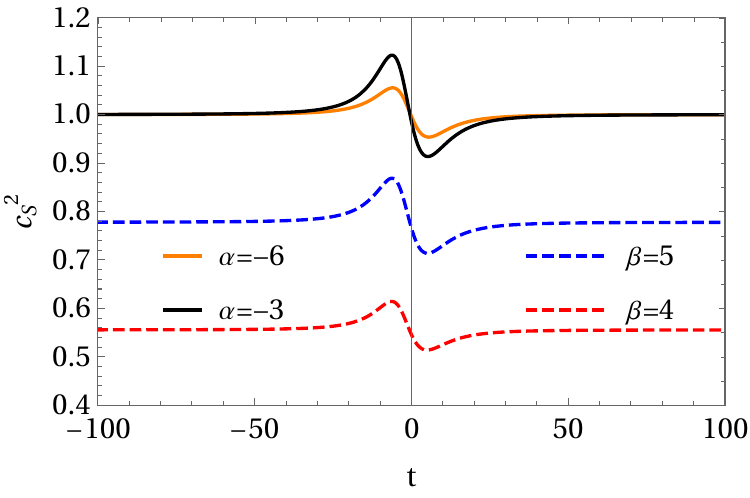} 
    \includegraphics[width=0.32\textwidth]{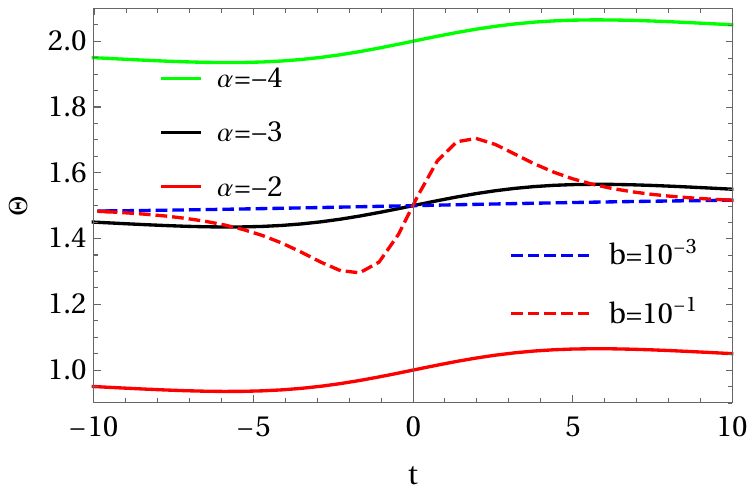}    \includegraphics[width=0.32\textwidth]{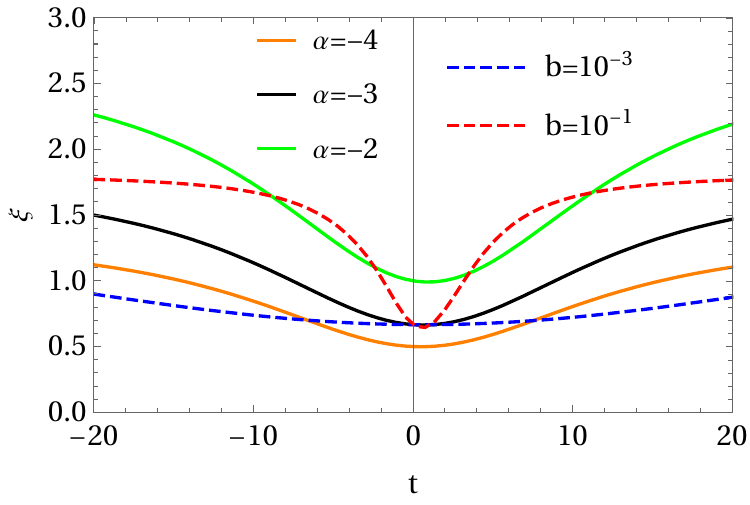}  
    \caption{Model III: The scalar perturbation coefficients are plotted for different values of $\alpha, \beta$ and $b$ parameters. }
    \label{fig: model III - scalar}
\end{figure}

In the last model considered, we assume a different bouncing scale factor, that reads
\begin{equation}\label{eq:a-bouncing2}
    a (t) = a_0 e^{\frac{b t^2}{1+ b t^2}}\,,
\end{equation}
while the scalar field remains the same, i.e. $\phi(t) = t$. The Horndeski functions take the form
\begin{align}
    G_2 &= - X^2\,\Bigg[\frac{4b \left( -2+3\alpha \phi -8b \phi^2+6b \alpha \phi^3 + 6b^2 \phi^4 +3b^2 \alpha \phi^5 \right)}{(1+b\,\phi^2)^4}\Bigg]\,,\\
    G_{3}&=\frac{b\phi \left(9 + 20 b \phi^2 +3 b^2 \phi^4 \right)}{2(1+b\,\phi^2)^3}+\frac{3\, \alpha}{2(1+b\,\phi^2)}+\frac{3\sqrt{b}}{2}\,\rm tan^{-1}(\sqrt{b}\phi) \,,\\
    G_4&=\frac{1}{2}+X \,,\\
    G_5&=\phi \,,
\end{align}
and the teleparallel Lagrangian reads
\begin{equation}
    G_{\rm Tele}=\alpha I_2+\beta\,J_3\,.
\end{equation}
The tensor perturbation coefficients in this model are always equal to unity,
\begin{align}\label{example3}
    \mathcal{G}_{\rm T}=\mathcal{F}_{\rm T}=1 \,, 
\end{align}
thus having always a constant speed of gravitational waves that is equal to one. In addition, the scalar perturbation coefficients take the form
\begin{gather}
    \mathcal{A} = 1\,,\,\,\,\mathcal{B}=1-\frac{2\beta}{3} \,,\,\,\,\mathcal{C}=1+\frac{\beta}{6} \nonumber \\ 
    \Sigma=9\,H^2-\frac{9 \alpha}{2}\,H+3\dot{H}\,,\,\,\, \Theta=H-\frac{\alpha}{2}\,,\\
    \mathcal{G}_{\rm S}=3\,\Bigg(1+2\frac{6H^2+2\dot{H}-3\alpha H}{(2H-\alpha)^2}\Bigg) \,,\,\,\, \mathcal{F}_{\rm S}=2\frac{2H^2-\alpha H-2\dot{H}}{(2H-\alpha)^2}(1+\frac{\beta}{6})-\frac{1}{(2H-\alpha)^2}(1-\frac{2\beta}{3})\,.
\end{gather} 
In Fig.~\ref{fig: model III - scalar} we plot these quantities for different values of $\alpha, \beta$ and $b$. Specifically, for $\mathcal{F}_{\rm S}$ and thus $c_{\rm S}^2$, the value of $\beta$ changes the height of the plot, while the value of $\alpha$ distorts its shape. Furthermore, the $\Theta$ plot changes its height for different values of $\alpha$, while $b$ changes its shape; however, it remains positive and non-vanishing everywhere, and thus the solution stable. 

\section{Conclusions} \label{sec:conclusion}

The no-go theorem is central to the study of extremely early Universe physics related to the singularity problem in the concordance model of cosmology. This is important because classical solutions to this problem are mainly concerned with continuous solutions at and near the point of vanishing cosmic time. While quantum theory is expected to dominate the dynamics of matter in this era, modifications to the gravitational sector of the standard cosmological model may avoid the need for this sector to break down entirely. In the regular formulation of Horndeski gravity \cite{Horndeski:1974wa}, a single scalar field is coupled to the Levi-Civita connection with the full spectrum of possible realizations of the theory included. As explained in detail in Sec.~\ref{sec:BDLS_intro}, once stability and strong coupling conditions are taken into consideration, it is not possible to find solutions for the curvature-based Horndeski theory that has non-singular solutions that resolve the singularity problem. This is the essence of the no-go theorem. The topic has become even more constrained in recent years with the restrictions on the Horndeksi action coming from observations of gravitational waves with strict limits to the speed of propagation of these waves.

By considering an exchange of curvature with torsion through the teleparallel connection, the BDLS adaptation of Horndeski can be constructed. The most meaningful difference here is the exchange of geometric connections, otherwise, the construction remains unchanged. Given the lower-order nature of TG, the action turns out to contain the regular Horndeski Lagrangian together with a new term~\eqref{action}. It is this new term that opens the possibility of evading the no-go theorem, as expressed in Eq.~\eqref{assume1}. The prospect of producing healthy non-singular cosmological solutions for BDLS theory would be a clear asset for the class of models. On the other hand, there is no guarantee of this possibility and no-go theorem evading solutions may only occur in some circumstances due to the inequality formulation of the expression in question.

In this work, we present four realizations of the BDLS Lagrangian terms that offer the possibility of evading the no-go theorem, while continuously satisfying the speed of gravitational waves constraint. In all these cases, the $G_4$ and $G_5$ contributions are non-trivial. In the first~(\ref{eq:model1_G2}--\ref{eq:model1_GT}) and second~(\ref{eq:model1_G2}--\ref{eq:model1_GT}) models are based on the same solution for the scale factor expansion profile and scalar field solution~\eqref{eq:a-bouncing}, which take on the form of a power law model and linear cosmic time factor respectively. In the first model, we explore an interesting scenario in which a simple coupling term between the torsion scalar and the kinetic term together with a $J_5$ linear term, which vanishes at the background level. The scalar perturbation propagation speed observes an antisymmetric profile about the zero cosmic time point, while being always positive for a large range of model parameter values, as shown. The profile width and amplitude can be adjusted through the two model parameters, which gives a fairly versatile counter example in which the no-go theorem is clearly evaded in a healthy manner.

The second model explored here has a different construction and is formulated as a linear function of both the torsion scalar and the $J_3$ term. The standard Horndeski terms are largely the same as the first model except for the $G_4$ and $G_5$ terms which are slightly different. As in the first case, the speed of propagation of gravitational waves remains unity for all time. However, the speed of propagation of scalar perturbations again has an antisymmetric form about the zero cosmic time point and can be stretched or shifted depending on the values of the two model parameters. The result is a vast range of values for which this behavior indicates a healthy branch for violating the no-go theorem. The underlying behavior here may be sourced by the form of the scale factor, but this shows that within this class of models, healthy scenarios can be found. Finally, the third model scenario being shown here is based on an exponential form of the scale factor, but in this case, the cosmic time parameter is squared so that it will be symmetric about the initial zero point, and non-singularity overall. A combination of regular Horndeski terms together with a linear $G_{\rm Tele}$ contribution (on the $I_2$ and $J_3$ terms) is adopted. The end result is again a well-behaved scalar perturbations propagation speed.

In both scale factor cases, the initial singularity problem is averted since the cosmology is non-singular at that point. It would be interesting to investigate other possible scale factors that may give this property. Furthermore, possible future work could include a study of how these different very early Universe scenarios may affect the power spectrum of the cosmic microwave background radiation.

\section*{Acknowledgements}\label{sec:acknowledgements}
This article is based upon work from COST Action CA21136 Addressing observational tensions in cosmology with systematics and fundamental physics (CosmoVerse) supported by COST (European Cooperation in Science and Technology). JLS would also like to acknowledge funding from ``The Malta Council for Science and Technology'' as part of the REP-2023-019 (CosmoLearn) Project. KFD's work was supported by the PNRR-III-C9-2022–I9 call, with project number 760016/27.01.2023. 
BA acknowledges support by grants F-FA-2021-432, and MRB-2021-527 from the Agency for Innovative Development of Uzbekistan.

\appendix
\section{Analytic expressions of perturbative quantities} \label{Sec:appendix}
The analytic expression of $\Sigma$ and $\Theta$ in Eq.~\eqref{scalarcoef1},\eqref{scalarcoef2} are 
\begin{align}\label{eq:Sigma}
    \Sigma:=& X \left(G_{{\rm Tele},X}+2 X G_{{\rm Tele},XX}+2 X G_{2,XX}+G_{2,X}-2 X G_{3,\phi X}-2 G_{3,\phi}\right)+3 H \dot{\phi} \big(4 X G_{{\rm Tele},X I_{2}}+G_{{\rm Tele},I_{2}} \nonumber \\ 
   &+2 X^2 G_{3,XX}+4 X G_{3,X}-4 X^2 G_{4,\phi X X}-10 X G_{4,\phi X}-2 G_{4,\phi}\big)+3 H^2 \big(12 X G_{{\rm Tele},I_{2} I_{2}}+8 X G_{{\rm Tele},X T} \nonumber \\
   &+2 G_{{\rm Tele},T}-2 G_{4}-3 G_{{\rm Tele},T_{\rm vec}}+8 X^3 G_{4,XXX}+32 X^2 G_{4,XX}+14 X G_{4,X}-12 X G_{{\rm Tele},XT_{\rm vec}}-4 X^3 G_{5,\phi XX} \nonumber \\ 
   &-18 X^2 G_{5,\phi X}-12 X G_{5,\phi} \big)+2 H^3 \dot{\phi} \left(36 G_{{\rm Tele},TI_{2}}-54 G_{{\rm Tele},T_{\rm vec} I_{2}}+2 X^3 G_{5,XXX}+13 X^2 G_{5,XX}+15 X G_{5,X}\right) \nonumber \\ 
   &+18 H^4 \left(-12 G_{{\rm Tele},TT_{\rm vec}}+4 G_{{\rm Tele},TT}+9 G_{{\rm Tele},T_{\rm vec} T_{\rm vec}}\right), \\ \label{eq:Theta}
    \Theta:=& -6 H^3 (4 G_{{\rm Tele},TT}-12 G_{{\rm Tele},TT_{\rm vec}}+9 G_{{\rm Tele},T_{\rm vec}T_{\rm vec}})+ H(2 G_{4}-8 X G_{4,X}-8 X^2 G_{4,XX}+6 X G_{5,\phi}\nonumber \\ 
    &+4 X^2 G_{5,\phi X}-2 G_{{\rm Tele},T}+3 G_{{\rm Tele},T_{\rm vec}}-6 X G_{{\rm Tele},I_{2}I_{2}}-4 X G_{{\rm Tele},XT}+6 X G_{{\rm Tele},XT_{\rm vec}}) \nonumber \\ 
    &- H^2 (5 X G_{5,X}+2 X^2 G_{5,XX}+18 G_{{\rm Tele},TI_{2}}-27 G_{{\rm Tele},T_{\rm vec}I_{2}}) \dot{\phi} \nonumber \\
    &-(X G_{3,X}-G_{4,\phi}-2 X G_{4,\phi X}+\frac{1}{2}G_{{\rm Tele},I_{2}}+ X G_{{\rm Tele},XI_{2}}) \dot{\phi} \, . 
\end{align}
The expressions for $f_i(G_{\rm Tele}$ with $i = (1,2,3)$ are
\begin{align}
    \label{f1}
    f_1\left(G_{\rm Tele}\right) &= 3G_{\rm Tele,T_{\rm vec}}-X\left(G_{\rm Tele,J_5}+4G_{\rm Tele,J_8}+3G_{\rm Tele,I_2 I_2}\right)+6H \dot{\phi}\left(3G_{\rm Tele,T_{\rm vec}I_2}-2G_{\rm Tele,TI_2}\right)+ \nonumber \\ 
    &\,\,\,\,+6H^2\left(12G_{\rm Tele,TT_{\rm vec}}-4G_{\rm Tele,TT}-9G_{\rm Tele,T_{\rm vec}T_{\rm vec}}\right),
    \\ 
    \label{f2}
    f_2\left(G_{\rm Tele}\right)&=\frac{2}{9}\left(18G_{\rm Tele,T_{\rm vec}}-6XG_{\rm Tele,J_3}-5XG_{\rm Tele,J_5}-2XG_{\rm Tele,J_8}+2X^2G_{\rm Tele,J_6}\right),
    \\ 
    \label{f3}
     f_3\left(G_{\rm Tele}\right)&=\frac{1}{9}\left(18G_{\rm Tele,T_{\rm vec}}+3XG_{\rm Tele,J_3}+XG_{\rm Tele,J_5}-32XG_{\rm Tele,J_8}-4X^2G_{\rm Tele,J_6}  \right).
\end{align}

\bibliography{references}

\end{document}